# Thermoelectric signatures of the electron-phonon fluid in PtSn$_4$

Chenguang Fu[1], Thomas Scaffidi[2], Jonah Waissman[4], Yan Sun[1], Rana Saha[5], Sarah J. Watzman[6], Abhay K. Srivastava[5], Guowei Li[1], Walter Schnelle[1], Peter Werner[5], Machteld E. Kamminga[7], Subir Sachdev[4], Stuart S. P. Parkin[5], Sean A. Hartnoll[3], Claudia Felser[1], Johannes Gooth[1,4*]

*[1]Max Planck Institute for Chemical Physics of Solids, 01187 Dresden, Germany.*

*[2]Department of Physics, University of California, Berkeley, CA 94720, USA.*

*[3]Department of Physics, Stanford University, Stanford, CA 94305-4060, USA.*

*[4]Department of Physics, Harvard University, Cambridge, MA 02138, USA.*

*[5]Max Planck Institute of Microstructure Physics, 06120 Halle, Germany.*

[6]Department of Mechanical and Aerospace Engineering, The Ohio State University, Columbus, OH 43210, USA

[7]Zernike Institute for Advanced Materials, University of Groningen, Nijenborgh 4, 9747 AG Groningen, The Netherlands.

\* *Correspondence to:* johannes.gooth@cpfs.mpg.de

**In most materials, transport can be described by the motion of distinct species of quasiparticles, such as electrons and phonons.[1] Strong interactions between quasiparticles, however, can lead to collective behaviour, including the possibility of viscous hydrodynamic flow.[2,3] In the case of electrons and phonons, an electron-phonon fluid is expected to exhibit strong phonon-drag transport signatures[4] and an anomalously low thermal conductivity.[5,6] The Dirac semi-metal $PtSn_4$ has a very low resistivity at low temperatures[7] and shows a pronounced phonon drag peak in the low temperature thermopower[7]; it is therefore an excellent candidate for hosting a hydrodynamic electron-phonon fluid. Here we report measurements of the temperature and magnetic field dependence of the longitudinal and Hall electrical resistivities, the thermopower and the thermal conductivity of $PtSn_4$. We confirm a phonon drag peak in the thermopower near 14 K and observe a concurrent breakdown of the Lorenz ratio below the Sommerfeld value. Both of these facts are expected for an electron-phonon fluid with a quasi-conserved total momentum[5,6,8]. A hierarchy between momentum-conserving and momentum-relaxing scattering timescales is corroborated through measurements of the magnetic field dependence of the electrical and Hall resistivity and of the thermal conductivity. These results show that $PtSn_4$ exhibits key features of hydrodynamic transport.**

Conventional transport in condensed matter systems relies on the existence of two species of quasi-particles, electrons and phonons.[1] Heat is carried by both electrons and phonons, while charge is only transported by electrons. The charge current can only be relaxed by processes that degrade the total momentum of the electrons. The simplest example is electron-impurity scattering which instantly releases momentum to the environment. The case of electron-phonon scattering is more subtle: in this case, the electron momentum is transferred to the phonons and could be transferred back to the electrons at a later scattering event. This kind of process is usually negligible because phonons quickly lose momentum to the environment (ex. *via* phonon-phonon Umklapp scattering), therefore acting effectively as a momentum sink. We consider instead the case where phonons lose their momentum to the environment more slowly than the electron-phonon scattering rate. In this case, electrons and phonons effectively form one single fluid since the rate of collision between them is the fastest process.[2] The electron-phonon fluid can then have a long-lived total momentum if impurity and Umklapp scattering are suppressed. This leads to a hydrodynamic description of transport in terms of a velocity field, including viscous effects due to the transverse diffusion of momentum density.[3]

Recently, signatures of hydrodynamic electronic transport have, in fact, been observed in graphene,[9–11] GaAs,[12] $PdCoO_2$[13] and $WP_2$.[14] Thermal and electrical transport experiments in these materials and others have revealed a variety of electron hydrodynamic properties, such as a size-dependent electrical resistance,[12–14] the violation of the Wiedemann-Franz law,[9,14,15] super-ballistic flow,[11] and the formation of whirlpools in non-local measurements.[10] Moreover, signatures of hydrodynamic phonon transport have been observed for phonon systems as the second sound phenomenon in solid helium[16] and have recently been shown to play a major role in the thermal transport of graphite.[17] This article departs from previous work by studying a system in which the strongly coupled electron-phonon fluid exhibits features characteristic of a hydrodynamic regime. Interest in experimental signatures of electron-phonon hydrodynamics has been accelerated by the recent proposal that hydrodynamic effects play a role in the *T*-linear resistivity of unconventional metals such as high-temperature superconductors above their critical temperature.[18–20] It is therefore essential to establish key signatures of hydrodynamic flow in strongly interacting electron-phonon systems.

To address the hydrodynamic electron-phonon fluid in a transport experiment, we sought to identify a material where momentum-relaxing scattering is supressed and electron-phonon interactions are strong. We have chosen the layered Dirac semimetal $PtSn_4$ with *a-c* planes stacked perpendicularly along the *b*-axis (Fig 1 (a)). Its electrical resistivity $\rho$ behaves like a metal with a Debye temperature of $\Theta_D = 210$ K. $\rho$ is remarkably low, with $\rho \approx 40$ nΩ and a mean free path $\lambda_{mr}$ of the order of microns at 2 K in the *a-c* plane of the crystal.[7] Despite the rather complex Fermiology (Extended Data Fig. 5), the effective masses of the electron $m_e^*$ and of the hole pockets $m_h^*$ are very similar ($m_{e/h}^* \approx (0.2 \pm 0.1)\, m_0$),[7] where $m_0$ is the free electron mass (see methods for details). In previous studies, such low electrical resistivity has been proven to be a good indicator for hydrodynamic electron systems, *e.g.* in $PdCoO_2$[13] and in $WP_2$.[14] $PtSn_4$ additionally exhibits a significant phonon-drag peak in the temperature-dependent thermopower $S$ at around 14 K,[7] indicating strong electron-phonon interactions. This makes the material an ideal candidate to search for signatures of the hydrodynamic electron-phonon fluid. For our experiments, we use $PtSn_4$ single crystals, grown out of a Sn-rich binary melt.[7] Respective growth-details and chemical and structural characterization can be found in the Methods section. The electrical resistivity $\rho$, the Hall resistivity $\rho_H$, the thermopower $S$, and the thermal conductivity $\kappa$ have been measured from 2 K to 300 K at magnetic fields $B$ up to 9 T. Details of the measurement procedure can also be found in the Methods section. All transport experiments are conducted within the *a-c* plane of the crystals and the magnetic field $B$ is applied along the *b*-axis.

In a first set of transport experiments at zero magnetic field, we establish that the $PtSn_4$ bulk crystals exhibit the low electrical resistivity $\rho$ previously observed. In Fig. 1 (b), the measured $\rho$ is shown as a function of $T$. In agreement with literature,[7] $\rho(T)$ at zero field increases linearly with $T$ between 25 K and 300 K, as expected for such semi-metals in which electron-phonon scattering is thought to dominate. Below 8 K the resistivity starts to saturate with a residual resistivity of 45 n$\Omega$cm at 2 K. Furthermore, a high residual resistance ratio of $\rho(300\text{ K})/\rho(2\text{ K}) \approx 1000$ is observed, which together with the non-saturating quadratic magneto-resistivity $MR = (\rho(B) - \rho(0\text{ T}))/\rho(0\text{ T})$ up to 9 T (Fig. 1 (c)) indicates a long $\lambda_{mr}$.

The transport properties of our $PtSn_4$ samples are further characterized by measurements of the Hall resistivity $\rho_H$ (Fig. 1 (d)) as a function of $B$ at various $T$. Upon cooling from 300 K, $\rho_H$ exhibits a positive slope as a function of $B$, indicating the that hole-like carriers dominate the transport. At around 50 K, the slope of $\rho_H(B)$ reverses its sign. Given that $d\rho_H(B)/dB$ changes sign at low temperatures, it is tempting to associate this phenomenon with transport contributions of both, electron and holes, which has previously been also invoked as an explanation for the non-saturated MR.[21] This scenario is supported by the non-linear behaviour of the $\rho_H(B)$ curves in high magnetic field. For a more quantitative analysis of the carrier densities, we calculated the Hall conductivity $\sigma_H = \rho_H / (\rho_H^2 + \rho^2)$ (Fig. 2 (e)) and employ a two-channel model:

$$\sigma_H(B) = \left[\frac{p\mu_h^2}{1+\mu_h^2 B^2} - \frac{n\mu_e^2}{1+\mu_e^2 B^2}\right] eB \qquad (1)$$

The model allows us to obtain the temperature-dependent average carrier density $n$, $p$ and the mobility $\mu_e$, $\mu_h$ for the electrons and hole pockets, respectively (see methods for details). Best fits to our data indeed reveal that our PtSn$_4$ samples exhibit parallel transport of both electrons and holes across the full temperature range investigated (Fig. 1 (f)), with a hole-excess above 50 K and an electron access below. The mobility of the two types of carriers are very similar (Fig. 1 (f), inset), indicating that scattering across the whole Fermi surface is dominated by a single scattering time.

Next, we explore whether coupling between the charge carriers and phonons plays a role in the transport properties of our PtSn$_4$ crystals. The thermopower $S$ is a sensitive probe for such coupling, since it is greatly enhanced by phonon drag, whereby a non-equilibrium distribution of phonons and electrons drift jointly. Such phonon drag appears if the momentum exchange between electrons and phonons is much faster than momentum-relaxing processes (such as phonon-phonon Umklapp).[1,22] Lowering the temperature leads to a suppression of Umklapp processes, but also to a lowering of electron-phonon scattering. This trade-off leads to a peak in the thermopower at intermediate temperatures, with a common theoretical estimate of the peak temperature $T_{PD}$ given by $\Theta_D/10$.[23] Consistent with the Hall data, $S(T)$ is positive at high

*T*, signifying that hole-type carriers dominate the thermoelectric transport (Fig. 2 (a)). Upon cooling, *S* exhibits a sign reversal at 50 K followed by a large, negative phonon-drag peak centred around $T_{PD}$ = 14 K, which is consistent with the literature.[7] The onset of the phonon-drag peak at around 22 K ≈ 0.1 $\Theta_D$ in our zero-field measurement data matches this theoretical estimate excellently and indicates the emergence of a strongly coupled electron-phonon system in PtSn$_4$.

This notion is supported by the magnetic field dependence of the thermopower at low temperatures (Fig. 2 (a)). While above 22 K, *S* depends only weakly on *B*, around the phonon-drag peak, however, the magneto-thermopower *MS* = (*S*(*B*) - *S*(0 T)) / *S*(0 T) becomes significant at higher fields (*B* > 3 T), following a linear relation in *B* (Fig. 2 (b)). The peak position remains located near $T_{PD}$ as the magnetic field increases. The linear relation at high magnetic fields can be understood by the increase of the electron-phonon scattering rate with *B*,[24] yielding $MS \approx \hbar\omega_c/k_B T$ for $\hbar\omega_c > k_B T$, where $\hbar$ and $k_B$ are the reduced Planck and the Boltzmann constants, respectively. $\omega_c = eB/m_{e/h}^*$ denotes the cyclotron frequency with the elementary charge *e*. Employing the average effective mass $m_{e/h}^* = 0.2\ m_0$, the linear relation between *MS* and *B* at above 3 T is expected for T < 22 K, which matches the experiments excellently. In hydrodynamics, the increasing electron-phonon coupling leads to a greater entropy per charge carrier in the electron-phonon fluid, which increases the magnitude of the thermopower. This *B*-dependence of the thermopower is weaker than that of the low temperature resistivity (Fig. 1 (c)), which varies over several orders of magnitude.

To determine whether this coupled electron-phonon state exhibits fluid characteristics, we explore the Lorenz ratio $L = \kappa_e/\sigma T$, where $\kappa_e$ is the electronic thermal conductivity and $\sigma$ the electrical conductivity. The Wiedemann-Franz (WF) law states that the Lorenz ratio is a constant given by the Sommerfeld value $L_0 = 2.44\times 10^{-8}$ W$\Omega$K$^{-2}$.[1] Central to this statement is that scattering affects the relaxation of both charge and heat currents in the same way. As a robust prediction of Fermi liquid transport theory, the WF law has been verified in numerous metals.[1,22] A breakdown of the Lorenz ratio $L/L_0 < 1$ is typically an indication of inelastic electron-phonon scattering[22] or unconventional phases of matter, such as in incoherent diffusive metals,[15] Luttinger liquids,[25] and metallic ferromagnets[26]. For finite-electron density hydrodynamical systems, $L/L_0$ can become arbitrarily small: The electrical conductivity is large due to slow momentum-relaxing processes, whereas the thermal conductivity is instead dominated by faster momentum-conserving collisions. This happens because the momentum-relaxing contribution to the thermal conductivity is proportional to the entropy density that is small at low temperatures[9] and furthermore the measurement of $\kappa$ with open circuit boundary conditions projects out the long-lived momentum mode.[5]

Thus, we measured the thermal conductivity $\kappa$ of our samples as a function of temperature (see Fig. 2 (c)). The experiments were performed with open electrical contacts, prohibiting electric current flow. $\kappa(T)$ exhibits the $T$-dependence of a metal, consistent with measurements of $\rho$ and $S$. Upon warming from 2 K, $\kappa(T)$ linearly increases with $T$ due to dominant impurity scattering. Near 8 K, $\kappa(T)$ reaches a maximum and then starts to decreases as $\kappa_{e,0}T^{-2}$, obeying the Debye model[1] until it saturates above $\Theta_D$ at a $T$-independent value of $\kappa_{e,s} = 25$ Wm$^{-1}$K$^{-1}$. The pre-factor $\kappa_{e,0}$ includes band structure details.

Using $\sigma = 1/\rho$ and the total thermal conductivity $\kappa$, we calculate the Lorenz number $L = \kappa\rho/T$[9,15,27,28] (see Fig. 2 (d)) at zero magnetic field. At temperatures above 30 K and below 7 K, where phonon-phonon Umklapp and impurity scattering dominate the transport, respectively, no Wiedemann-Franz law violation can be established, ($L/L_0 > 1$), as expected. The ratio is consistent with an excess phonon contribution to the thermal conductivity. However, in the intermediate range between 7 K and 30 K, where phonon-drag is dominant, we find $L/L_0 < 1$. Remarkably, despite the full phonon contribution to thermal conduction being included in $\kappa$, the WF law is not obeyed, resulting in an upper bound on $L/L_0$ of 0.6 at $T_{PD} = 14$ K. The phonon drag-peak and the concurrent observation of $L/L_0 < 1$ at 14 K together give clear signatures of the presence of an electron-phonon fluid in PtSn$_4$.

To gain more quantitative information on the electron-phonon fluid in PtSn$_4$, we extract the mean free paths associated with charge currents ($\lambda_{mr}$) and thermal currents ($\lambda_{th}$), based on the extracted values of $\mu_{e/h}$ and on the measured $\kappa(B)$. To obtain the momentum-relaxing mean free path $\lambda_{mr,e/h}$ of the electron and holes, we use the expression of the mobility $\mu_{e/h} = ev_{F,e/h}\lambda_{mr,e/h} / m_{e/h}^*$, with the Fermi velocity $v_{F,e/h}$ and the effective mass $m_{e/h}^*$. Not only are the effective masses of the electron and hole pockets in PtSn$_4$ very similar, but the Fermi velocities of the individual pockets are too. Therefore, the average Fermi velocity $v_F = (4 \pm 1) \times 10^5$ m/s is used for the extraction of $\lambda_{mr,e/h}$ (see methods for details), yielding similar $\lambda_{mr}(T)$ for the electron and hole pockets of up to 3.2 µm at 2 K. In accordance with the small deviations of $v_F$ and $\lambda_{mr}(T)$ for all Fermi pockets,[7] the MR obeys Kohler's rule $MR = F[B/\rho(0\ T)]$, approximately proportional to $B^2$ (Fig. 2 (e)). This implies that the momentum-relaxing scattering time is the same at all points on the Fermi surface and therefore $MR \propto (\lambda_{mr}B)^2$. In a hydrodynamic regime, the momentum conserving timescale is faster than the momentum relaxing timescale, but this is irrelevant for Kohler's rule because the resistivity is dominated by momentum-relaxing scattering alone. The long mean free path $\lambda_{mr}$, combined with Kohler's rule, leads to the strong magnetoresistance observed in this material as well as other hydrodynamic materials.[21,29] We note, however, that the Fermi surface of PtSn$_4$ consists of multiple small pockets (Extended Data Figure 5) and that the above analysis gives only a rough estimate of $\lambda_{mr}$.

The magnetic field-dependent thermal conductivity[30] (Fig. 3 (a)) allows for a direct extraction of the relaxation length of the thermal current $\lambda_{th}$, using the expression: $\kappa = \kappa_0 + \kappa_{B0}/(1+(\omega_c\lambda_{th}/v_F)^2)$.[1,22] For conventional semimetals, this method has previously been very successful in separating the $B$-independent phonon contribution $\kappa_0 = \kappa_{ph}$ from the $B$-dependent electron contribution $\kappa_{el} = \kappa_{B0}/(1+(\omega_c \lambda_{th}/v_F)^2)$, where $\kappa_{B0}$ is the value at zero-field. By fitting our data to this formula, we extract $\lambda_{th}$ as a function of $T$ and compare it to $\lambda_{mr}(T)$ in Fig. 3 (b). Above 100 K, we find that $\lambda_{th}/\lambda_{mr} \approx 1$, and that the dominant electron-phonon scattering processes are highly effective at relaxing currents, as both scattering lengths are tied to the relaxation length scale $\lambda_{th} \approx \lambda_{mr} \approx \lambda_{planck} = v_F\hbar/(k_B T)$ associated with the quantum limit of dissipation.[31] Below 100 K, however, the relaxation of currents becomes less effective and $\lambda_{mr}$ ,$\lambda_{planck} > \lambda_{planck}$. Concurrently, $\lambda_{th}$ becomes shorter than $\lambda_{mr}$, exhibiting their largest deviation at around $T_{PD}$ with a minimum of $\lambda_{th}/\lambda_{mr} \approx 0.4$ (see inset in Fig. 3 (b)). In the limit of $T \to 0$ K, the scattering time ratio recovers the high-temperature value $\lambda_{th}/\lambda_{mr} \approx 1$. The observed $T$-dependence of the $\lambda_{th}/\lambda_{mr}$ ratio is in full agreement with the $T$-dependence of $L$, confirming our analysis and interpretation above.

In the electron-phonon fluid regime, $L$ is expected to increase with magnetic field strength, as the magnetic field dependence of $(\lambda_{th}B)^2$ is weaker than $(\lambda_{mr}B)^2$, providing an important cross-check for our results. We therefore calculate the Magneto-Lorenz number $ML = (L(B) - L(B = 0\text{ T}))/L(B = 0\text{ T})$ at various temperatures, using the measured $\kappa(B)$ and longitudinal component of the electrical conductivity tensor $\sigma(B) = \rho(B)/(\rho(B)^2+\rho_H(B)^2)$. In agreement with the electron-phonon fluid picture, we find the maximum $ML$ around $T_{PD}$ as shown in Fig. 3 (c). The linear increase of $ML$ with $B$ near $T_D$ is consistent with the linear increase of $MS$ with $B$. That is, the increasing electron-phonon coupling is linearly increasing the entropy of the fluid by dragging more phonons.

As explained above, $\lambda_{th}$ is reduced in comparison to $\lambda_{mr}$ at intermediate $T$ due to the presence of scattering processes which conserve the total momentum of the electron-phonon fluid but relax thermal currents. As such, to a first approximation the scattering length associated with these momentum-conserving processes ($\lambda_{mc}$) can be obtained by using $1/\lambda_{th} = 1/\lambda_{mc} + 1/\lambda_{mr}$. This leads to $\lambda_{mc} < \lambda_{mr}$ at around $T_{PD}$ with a minimum of $\lambda_{mc} \approx 0.6 \lambda_{mr}$. These results are consistent with the emergence of an electron-phonon fluid, where momentum-conserving scattering provides the smallest length scale in the system.

The shear viscosity in such systems is expected to be[3] $\eta = v_F \lambda_{mc}/5$, leading to an estimate of $\eta \approx 8 \times 10^{-2}$ m$^2$s$^{-1}$ at the mimimum $\lambda_{th}/\lambda_{mr}$-ratio. This value is of the same order as the kinematic shear viscosities measured in graphene,[10,32] PdCoO$_2$[13] and WP$_2$.[14] However, we emphasise that $\lambda_{mc}$ is only 40 % smaller than $\lambda_{mr}$, which means the system is roughly between a purely hydrodynamic ($\lambda_{mc} \ll \lambda_{mr}$) and purely Ohmic ($\lambda_{mc} \gg \lambda_{mr}$) regime, which becomes important when analysing results in restricted geometries, such as channel flow-resistance[33,34] and non-local- probe experiments. The observation of such classic signatures of viscous electron flow in this material may also be complicated by the presence of potential long-lived charge imbalance modes due to the presence of both electron and hole pockets.[20,35]

Our work provides a foundation for the study of electron-phonon fluid hydrodynamical transport in PtSn4 and other materials. An interesting candidate is PdCoO$_2$, a material know to exhibit viscous electrical flow resistance in narrow channels,[13] where thermoelectric signatures[36] similar to those observed for PtSn$_4$ were independently measured. In general, every material that shows phonon-drag is a potential host of an electron-phonon fluid if momentum-relaxation can be sufficiently suppressed. Much of our analysis – in particular of the field dependence of the electrical resistivity, thermal conductivity and the Hall resistivity– relies on the existence of well-defined quasiparticles and a corresponding small $\rho$. Thus, the electron-phonon fluid in PtSn$_4$ is distinct from the recently observed incoherent, highly resistive, electron-phonon soup in underdoped YBa$_2$Cu$_3$O$_{6+x}$.[37] Nonetheless a hydrodynamic analysis may pertain for such case also.[20]

In conclusion, we observed a phonon-drag-peak in the *T*-dependent thermopower *S* that arises concurrently with the breakdown of the WF law near 14 K in the Dirac semimetal PtSn$_4$, despite the full phonon contribution to the thermal conductivity being included. These coincident observations give evidence for the existence of an electron-phonon fluid in PtSn$_4$. Magneto-transport experiments allow the momentum-relaxing and thermal mean free paths to be extracted, corroborating the hydrodynamic picture and leading to an estimate of the viscosity of the electron-phonon fluid.

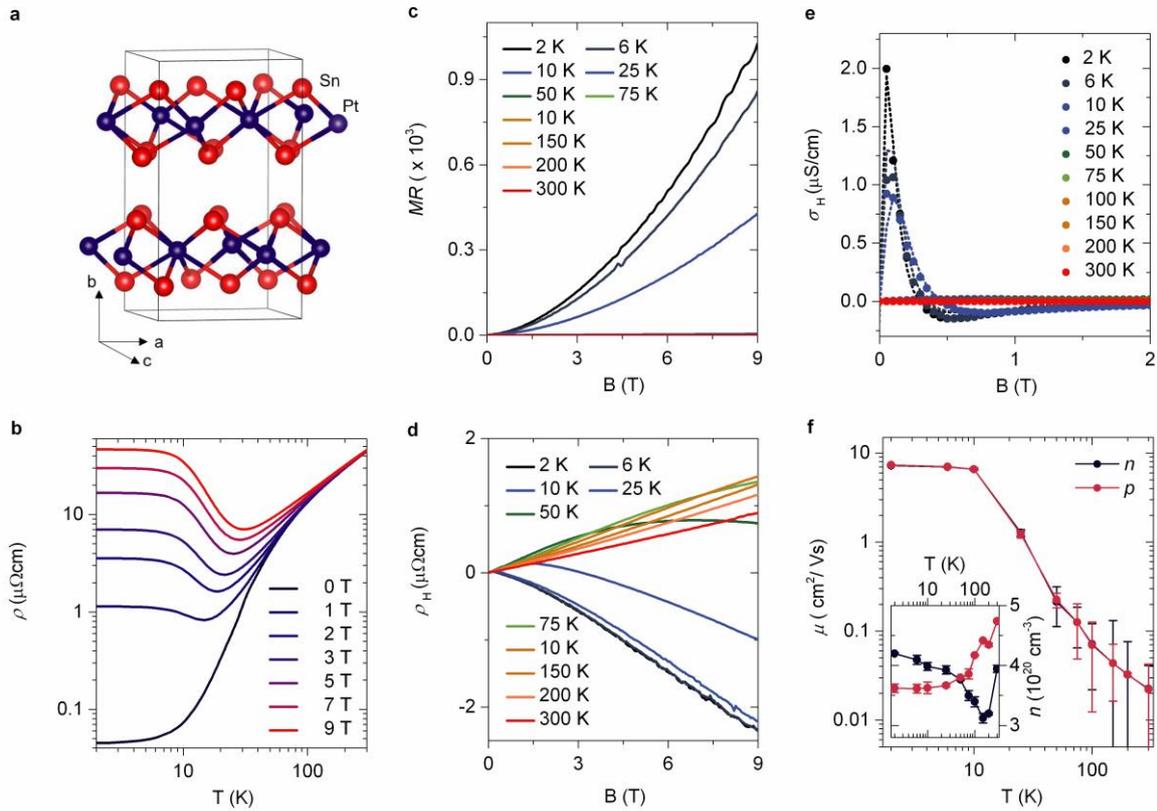

**Figure 1 | Electrical transport in a PtSn$_4$ single crystal. a**, Crystal structure of PtSn$_4$ at 100 K. The chemical bonds are drawn to denote the 8-fold Sn-coordinated Pt atoms. **b**, The electrical resistivity $\rho$ as a function of temperature $T$ at various magnetic fields $B$. **c.** Magentoresistivity $|MR| = |(\rho(B) - \rho(0\ \text{T})) / \rho(0\ \text{T})|$ and **d,** Hall resistivity $\rho_H(B)$ at various $T$. **e**, Hall conductivity $\sigma_H = \rho_H / (\rho_H^2 + \rho^2)$ at various temperatures. The symbols denote the measurement data. The dotted lines are fits according to a two-carrier model. **f**, Carrier concentration $n$, $p$ and mobility $\mu_e$, $\mu_h$ as a function of $T$ for the electrons and holes, respectively. The lines are a guide to the eyes. The error bars represent the error from the fits.

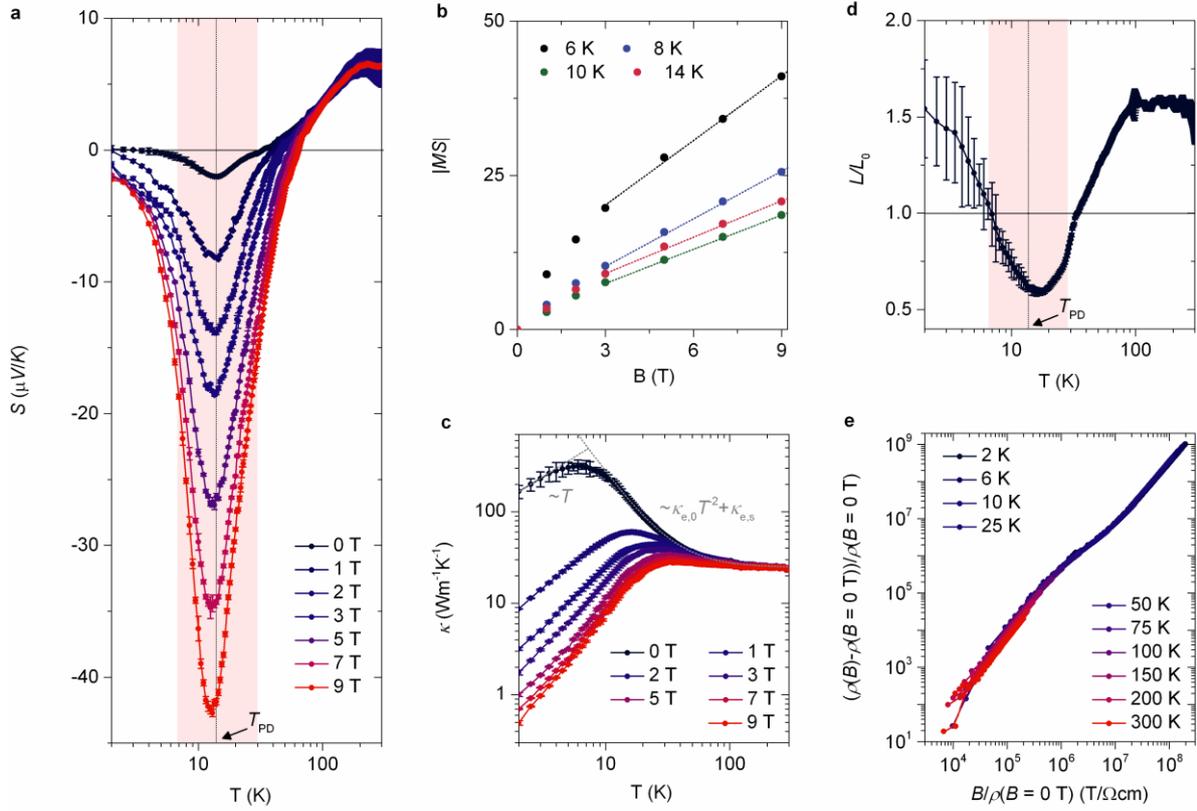

**Figure 2 | Thermoelectric and thermal transport in a PtSn$_4$ single crystal. a**, The thermopower $S$ as a function of $T$ at various $B$. The error bars represent the measurement error. At around $T_{PD} \approx 14$ K (marked by the dotted line) a phonon-drag peak evolves in $S$. Lines are guides to the eyes. The light red area marks the region, where the Wiedemann-Franz law is violated. **b,** Absolute magneto-thermopower $|MS| = |(S(B) - S(0\ T))/S(0\ T)|$ at various $T$. The data is extracted from (a). At high fields ($B > 3$ T), $|MS|$ can be well represented by linear fits (dotted lines), as expected by phonon-drag theory. **c**, The total thermal conductivity $\kappa$ as a function of $T$ at various magnetic fields $B$. **d**, The Lorenz ratio at zero magnetic field $L/L_0 = \kappa\rho/TL_0$ as a function of $T$, where $L_0 = 2.44\times10^{-8}$ W$\Omega$K$^{-2}$ is the Sommerfeld value. The error bars represent the measurement error. At around $T_{PD}$, $L/L_0$ violates the Wiedemann-Franz (indicated by light red region). **e,** The electrical resistivity obeys Kohler's rule $MR = F[B/\rho(0\ T)] \approx B^2$, indicating a single relaxation time at all points of the Fermi surface.

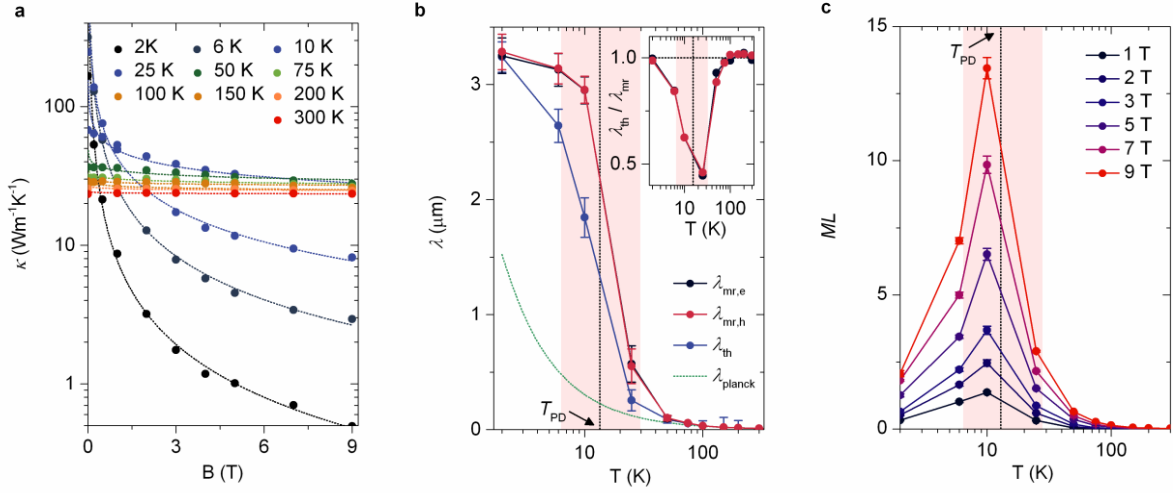

**Figure 3 | Relaxation length of the electrical $\lambda_{mr}$ and thermal current $\lambda_{th}$. a**, $\kappa$ as a function of $B$ for various $T$. The data (dots) is extracted from Fig. 2 (b). The dotted lines represent fits of $\kappa = \kappa_0 + \kappa_{B0}/(1+(\omega_c\lambda_{th}/v_F)^2)$ to the data, where $\kappa_0$ includes all phonons that are not coupled to the electron system, $\kappa_{B0}$ is the zero-field value of the $B$-dependent contribution of electrons and phonons and $\lambda_{th}$ is the relaxation length of the thermal currents. $\omega_c = eB/m^*$ is the cyclotron frequency with the elementary charge $e$, the effective mass $m^*$ and the Fermi velocity $v_F$. **b**, Relaxation length of the electrical $\lambda_{mr}$ and thermal current $\lambda_{th}$ as a function of $T$. $\lambda_{mr,e/h}$ is calculated using the mobility $\mu_{e/h} = ev_{F,e/h}\lambda_{mr,e/h} / m_{e/h}^*$. $\lambda_{th}$ is extracted from the fits of the data plotted in (a). The error bars are the errors obtained from the fits. $\lambda_{planck} = v_F \hbar/(k_B T)$ denotes the shortest possible relaxation length, sometimes called the "Planckian" limit of dissipation. **c**, Magneto-Lorenz number $|ML| = |(L(B) - L(0\,T)) / L(0\,T)|$ as a function of $T$ for various $B$.

## METHODS

### PtSn$_4$ single-crystal growth

Single Crystals of PtSn$_4$ were grown out of a Sn-rich binary melt, as described in the references 34 and 35. High-purity starting elements Pt (shot, 99.99%) and Sn (shot, 99.999%) were mixed together with an initial stoichiometry of Pt$_4$Sn$_{96}$, and afterward the mixture was placed in an alumina crucible and sealed in a quartz tube under a partial Ar pressure. The quartz tube was heated up to 600 °C over a period of 5h, and then kept for 20 h. After that, it was slowly cooled down to 350 °C over 60 h.[35] The excess Sn flux was removed by using a centrifuge at 400 °C. After the centrifugation process, the remaining flux on the surface was removed by mechanically polishing.

### Chemical and structural characterization of the PtSn$_4$ single-crystals

Powder X-ray diffraction measurement of PtSn$_4$ was performed with Cu K$\alpha$ radiation at room temperature to identify the phase purity and crystal structure, using an image-plate Huber G670 Guinier camera, diffraction range from $10° \leq 2\theta \leq 100°$ in step of 0.005°. Single crystal X-ray diffraction measurements were performed using a Bruker D8 Venture diffractometer equipped with a Triumph monochromator and a Photon100 area detector, operating with Mo K$\alpha$ radiation. The crystals were mounted on a 0.2 mm nylon loop using cryo-oil. The crystals were cooled with a nitrogen flow from an Oxford Cryosystems Cryostream Plus. Data processing was done using the Bruker Apex III software, the structures were solved using direct methods and the SHELX97 software[38] was used for structure refinement. A scanning electron microscope (SEM) with an attached energy-dispersive X-ray spectrometer (EDX) was used for elemental analysis.

The single crystal X-ray diffraction measurements of PtSn$_4$ shows that its crystal structure is described by the centrosymmetric space group *Ccca*, in agreement with previous reports.[39] Fig. 1 (a) shows the layered crystal structure of PtSn$_4$, consisting of PtSn$_4$ slabs that are constructed from 8-fold Sn-coordinated Pt atoms. The layered structure can be observed from the SEM results as shown Extended Data Fig. 1. Note that the layered nature of $PtSn_4$ allows for a slight disorder in the *ac*-plane, *i.e.* a small misalignment between consecutive $PtSn_4$ layers. Therefore, the overall quality of the single-crystal data is slightly compromised by the presence of aspherical spots. We have tested for twinning, but no twin model fitted our solution. Extended Data Table 1 shows the crystallographic and refinement parameters of $PtSn_4$. The highest residual peaks are relatively large (around 4 electrons at a distance of less than 1 Å from Pt), in close agreement with previous reports.[7] Extended Data Table 2 further shows the fractional atomic coordinates of the asymmetric unit and equivalent thermal displacement parameters.

Power X-ray diffraction pattern of $PtSn_4$ is shown Extended Data Fig. 2. The diffraction peaks can be well indexed to the orthorhombic structure, giving a good agreement with single crystal x-ray diffraction measurement. No obvious other phases are observed. The actual composition of $PtSn_4$ single crystal is determined by energy dispersive x-ray (EDX) spectroscopy at 7 randomly selected positions, which is in agreement with the nominal one, considering the instrument error, as shown in Extended Data Table S3.

High resolution transmission electron microscopy (HRTEM) (Extended Data Fig. 3) on a 6.7 μm × 4.4 μm large lamella showing the crystallinity of the $PtSn_4$ samples (Extended Data Fig. 4). However, dislocation lines with an area density of 0.07 $μm^{-2}$ were found.

**Thermoelectric transport measurements**

Temperature dependent thermal conductivity and thermopower under magnetic field were jointly measured by the one-heat and two-thermometer configuration using the thermal transport option (TTO) of the PPMS (Quantum Design) in which the sample was placed in an orientation where the magnetic field was perpendicular to the heat flow. The thermometers were calibrated under magnetic field by using the PPMS magneto-resistance calibration wizard before doing the thermal transport measurement. To make sure a uniform heat flow through the bar-shaped sample, two gold-plated copper leads were attached to the entire ends of the sample by using silver epoxy and then connected to the heater and sink, respectively. Another two copper leads were surrounded to the sample with silver epoxy and connected to the thermometers for detecting $\Delta V$ and $\Delta T$. The applied temperature gradients were around 1% - 3% of the base temperature. We now address the potential influence of thermal contact resistances. and demonstrated the necessity of a careful contact preparation. After careful contact preparation, as described above, we investigated samples with different length from 4.5 mm to 7 mm, keeping the sample cross-section and the contact area similar. We obtained the same results at all samples for any magnetic field or temperature investigated and therefore did not observe any indication of diminished thermal conductivity due to contact.

Both longitudinal and Hall resistivities were measured by a standard four-probe method using the AC transport option in a PPMS system with an AC current of 16 mA applied. Point contacts were made by spot-welding Platinum wires for Hall voltage probes. While for the current and longitudinal voltage probes, linear contacts were made by using silver paint and 25 μm Platinum wires. To correct for contact misalignment, the measured Hall resistivity was field anti-symmetrized. For all the transport measurements, the magnetic field was applied along *b*-axis, which is perpendicular to the *a-c* plane.

**Extraction of the transport parameters**

Employing fits of a two-channel model to the Hall conductivity $\sigma_H = \rho_H / (\rho_H^2 + \rho^2)$ (Fig. 2 (e)), we obtain the temperature (*T*)-dependent average carrier densities *n* for electron *n* and hole pockets *p* as well as the mobility for the electron $\mu_e$ and hole pockets $\mu_h$. For the fits we use the common expressions[20]

$$\sigma_H(B) = \left[\frac{p\mu_h^2}{1+\mu_h^2 B^2} - \frac{n\mu_e^2}{1+\mu_e^2 B^2}\right] eB \qquad (1)$$

The results are shown in Fig. 3 (c) and (d).

To obtain the momentum-relaxing mean free path $\lambda_{\text{mr,e/h}}$ of the electron and holes, we use the mobility $\mu_{\text{e/h}} = ev_{\text{F,e/h}}\lambda_{\text{mr,e/h}} / m_{\text{e/h}}^*$, with the Fermi velocity $v_{\text{F,e/h}}$ and the effective mass $m_{\text{e/h}}^*$. Despite the complex Fermiology, the effective masses of the electron and hole pockets in PtSn$_4$ are very similar as shown in Table 1 of reference 19. We therefore calculate the harmonic mean of all bands $i$, giving the average effective mass $m^* = i(\sum_i 1/m_i^*)^{-1} = (0.2 \pm 0.1)\, m_0$, where $m_0$ is the free electron mass. Using the Shubnikov-de Haas frequencies $f_i$ from the same table in reference 19, we determine the size of the Fermi surface cross sections from the Onsager relation $A_{\text{F}i} = 2\pi^2 f_i /\phi_0$ with $\phi_0$ as the magnetic flux quantum and apply the standard circular approximation to obtain the momentum-vectors $k_{\text{F}i} = (A_{\text{F}i}/\pi)^{1/2}$, from which the corresponding Fermi velocities can be calculated $v_{\text{F}i} = \hbar k_{\text{F}i}/m_i$, resulting in the average Fermi velocity $v_\text{F} = (\sum v_{\text{F}i})/i = (4 \pm 1)\times 10^5$ m/s. $\hbar$ is the reduced Planck constant.

## DATA AVAILABILITY STATEMENT

All data generated or analyzed during this study are available within the paper and its extended data files. Reasonable requests for further source data should be addressed to the corresponding author.

**EXTENDED DATA**

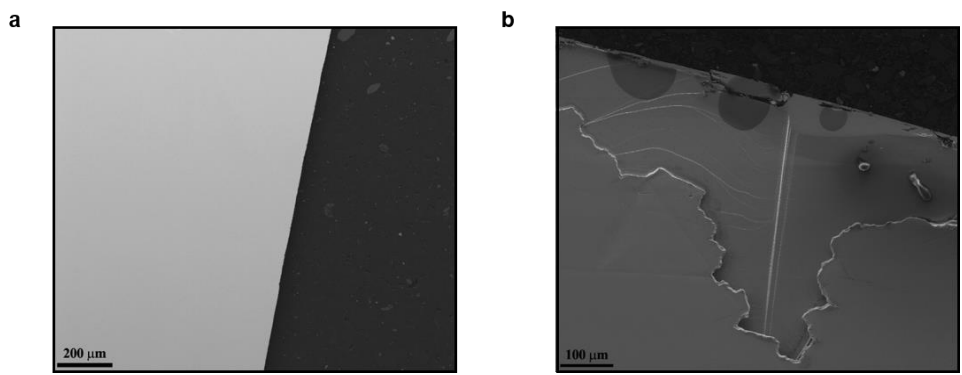

**Extended Data Figure 1 | a**. SEM back scattering and **b**. secondary electron images for the PtSn$_4$ single crystal.

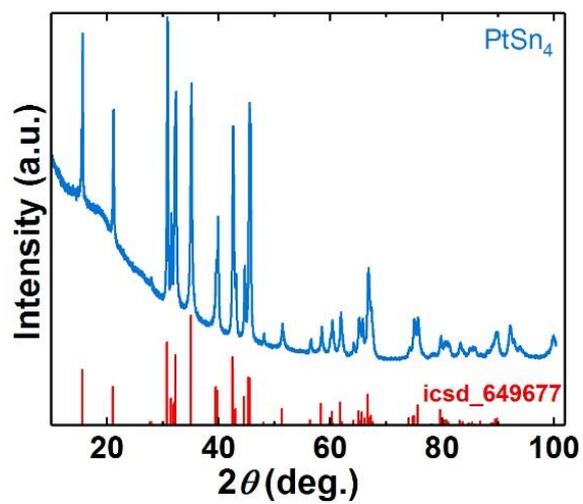

**Extended Data Figure 2 |** Power X-ray diffraction pattern of PtSn$_4$.

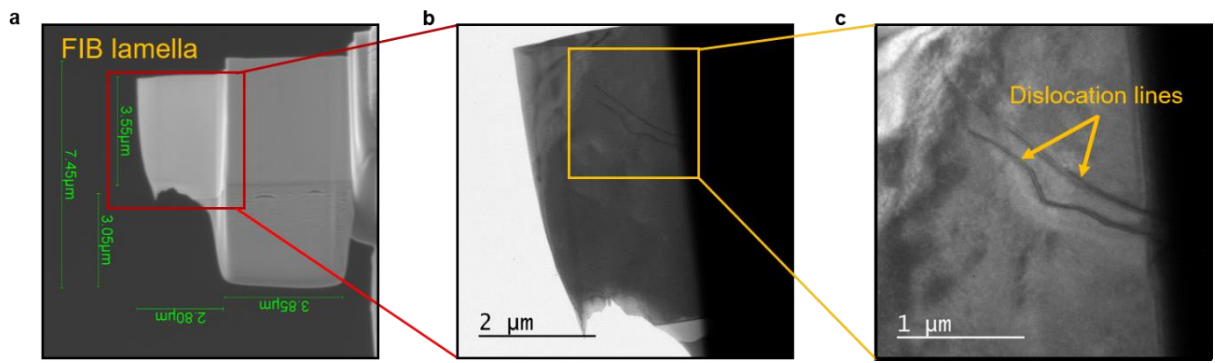

**Extended Data Figure 3 | Preparation of samples for transmission electron microscopy (TEM). a**. Scanning-transmission electron microscopy image showing the sample prepared by Focussed Ion Beam technique (FIB). The dimensions of the thin slice are inserted. **b**. Corresponding TEM micrograph of the thin area. "Diffraction contrast technique" was applied, which allows the identification of crystal defects. **c**. The magnified area shows to parallel dark lines, which were identified as dislocation.

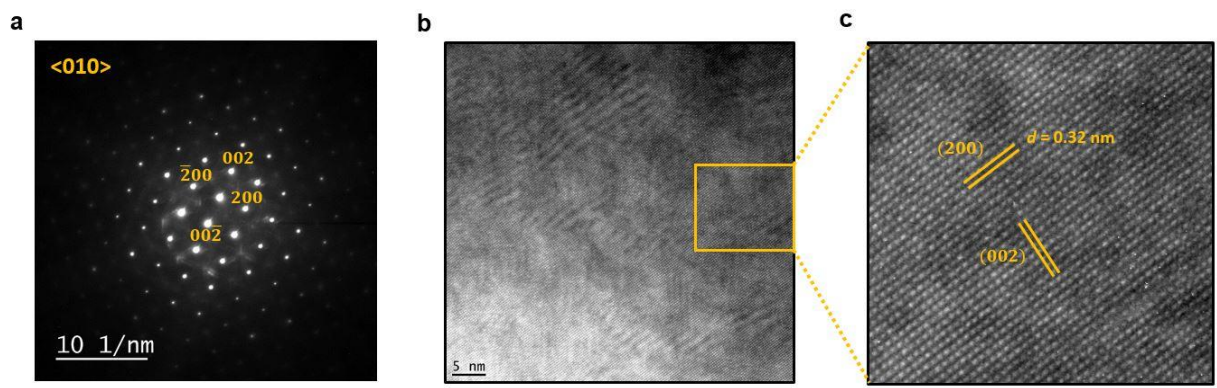

**Extended Data Figure 4 | Structure analysis by TEM. a**. Electron diffraction pattern received from the thin sample area. It correlates to a [010] crystal orientation. The corresponding reflections are indicated. **b**. TEM overview image. **c**. High-resolution TEM image showing the crystal lattice structure by {002} lattice planes with a distance $d$ of 0.32 nanometer.

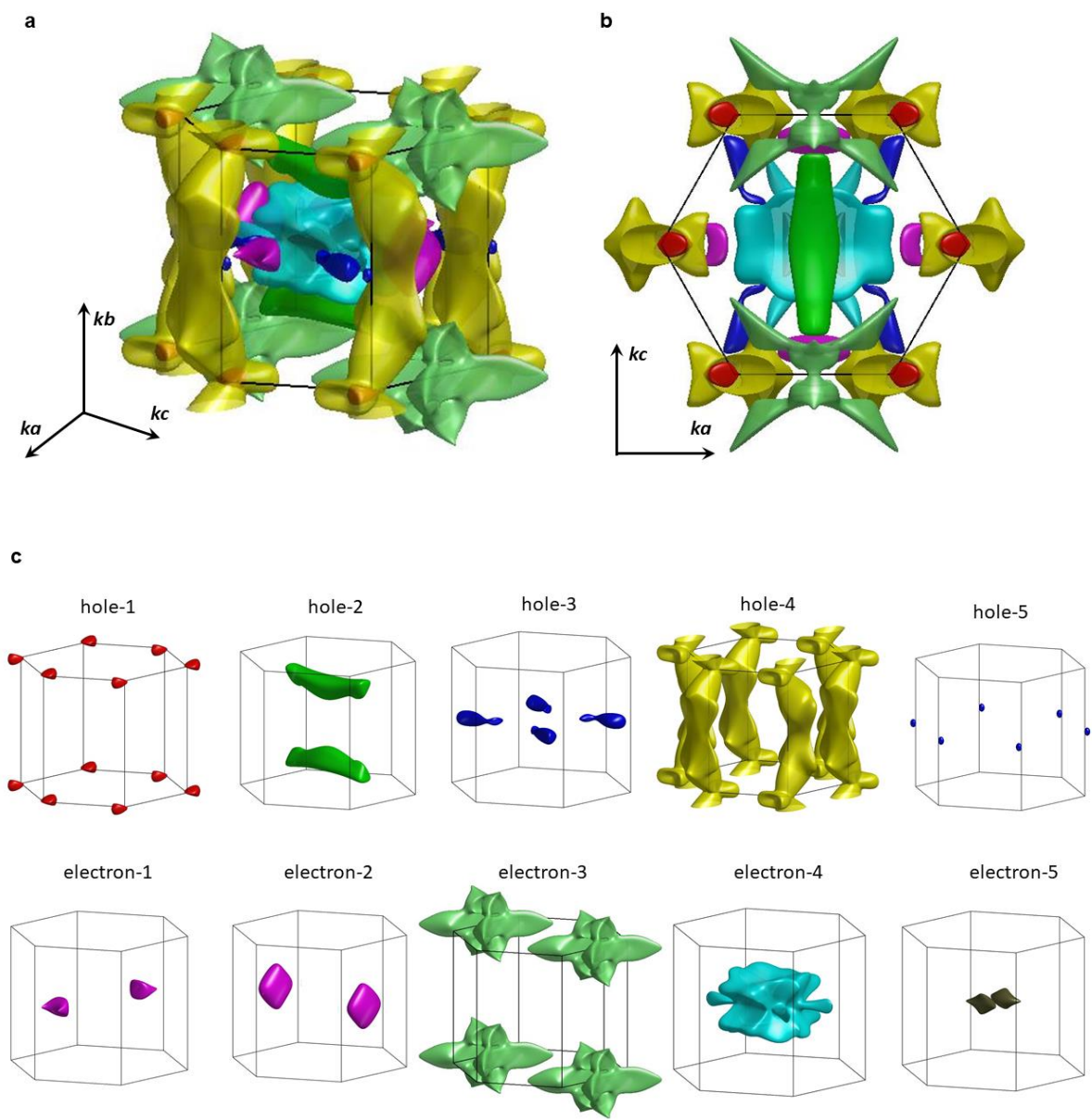

**Extended Data Figure 5 | Fermi surface of PtSn$_4$.** The Fermi surface of PtSn$_4$ is calculated using Density Functional Theory. **a**, Side view of the whole Fermi surface. **b**, Top view of the whole Fermi surface. **c**, Individual hole (upper row) and electron (lower row) pockets.

|  | PtSn$_4$ |
|---|---|
| **Crystal size** | 0.12 × 0.06 × 0.02 mm$^3$ |
| **Wavelength** | 0.71073 Å (Mo Kα radiation) |
| **Refinement method** | full matrix least squares on F$^2$, anisotropic displacement parameters |
| **Absorption correction** | multi-scan |

| crystal system | orthorhombic | min / max transmission factor | 0.1225 / 0.3706 |
|---|---|---|---|
| space group | *Ccca* (no. 68) | θ range (degrees) | 3.60 – 36.97 |
| symmetry | centrosymmetric | index ranges | -8 < h < 8 |
|  |  |  | -14 < k < 14 |
|  |  |  | -8 < l < 8 |
| Z | 4 | data / restraints / parameters | 287 / 0 / 13 |
| D (calculated) (g/cm3) | 9.652 | GooF on F$^2$ | 1.197 |
| F(000) | 556 | no. total reflections | 4675 |
| a (Å) | 6.4045(7) | no. unique reflections | 287 |
| b (Å) | 11.3087(15) | no. obs Fo > 4σ(Fo) | 240 |
| c (Å) | 6.3646(7) | R$_1$ [Fo > 4σ(Fo)] | 0.0289 |
| α (°) | 90.0 | R$_1$ [all data] | 0.0367 |
| β (°) | 90.0 | wR$_2$ [Fo > 4σ(Fo)] | 0.0634 |
| γ (°) | 90.0 | wR$_2$ [all data] | 0.0666 |
| volume (Å$^3$) | 460.97(9) | largest peak and hole (eÅ$^{-3}$) | 4.13 and -1.48 |
| absorption coefficient (mm$^{-1}$) | 51.405 |  |  |

**Extended Data Table 1 |** Crystallographic and refinement parameters of PtSn$_4$, measured at 100 K.

| Atom | x | y | z | U$_{eq}$ (Å$^2$) |
|---|---|---|---|---|
| Pt | 0.00000 | 0.25000 | 0.25000 | 0.0068(2) |
| Sn | 0.33436(13) | 0.12338(7) | 0.08470(13) | 0.0073(2) |

**Extended Data Table 2 |** Fractional atomic coordinates of the asymmetric unit and equivalent thermal displacement parameters.

|  | Position 1 | Position 2 | Position 3 | Position 4 | Position 5 | Position 6 | Position 7 | Average |
| --- | --- | --- | --- | --- | --- | --- | --- | --- |
| Pt | 22.11% | 21.58% | 21.89% | 21.00% | 22.18% | 20.87% | 21.16% | 21.54% |
| $Sn_4$ | 77.89% | 78.42% | 78.11% | 79.00% | 77.82% | 79.13% | 78.84% | 78.46% |

**Extended Data Table 3 |** Atomic percentage of PtSn$_4$ single crystal at 7 randomly selected positions detected by EDX.

## ACKNOWLEDGEMENTS

This work was financially supported by the ERC Advanced Grant No. (742068) "TOP-MAT". Ch.F. acknowledges financial support from the Alexander von Humboldt Foundation. M.E.K. was supported by The Netherlands Organization for Science NWO (Graduate program 2013, No. 022.005.006).

## AUTHOR INFORMATION

### Contributions

Ch.F., Cl.F., S.H., T.S. and J.G. conceived the experiment. Ch.F. synthesized the single-crystal bulk samples and carried out the thermoelectric transport measurements with the help of W.S. Y.S. did the DFT calculation. Ch.F., G.L. and M.E.K. performed SEM and single crystal XRD measurements. R.S., A.K.S., P.W. and S.P. prepared the FIB lamella and performed TEM measurements. T.S., S.H., J.W., S.S., S.J.W. and J.G. analyzed the data. J.G. supervised the project. All authors contributed to the interpretation of the data and to the writing of the manuscript.

### Competing financial interest

The authors declare no competing financial interests.

### Corresponding author

* johannes.gooth@cpfs.mpg.de